# FastTrack: a fast method to evaluate mass transport in solid leveraging universal machine learning interatomic potential

Hanwen Kang[1], Tenglong Lu[1], Zhanbin Qi[1,2],
Jiandong Guo[1]*, Sheng Meng[1,2]*, Miao Liu[1,2]*

[1]*Beijing National Laboratory for Condensed Matter Physics, Institute of Physics, Chinese Academy of Sciences, Beijing 100190, China*
[2]*Songshan Lake Materials Laboratory, Dongguan, Guangdong 523808, China*

*Corresponding author: jdguo@iphy.ac.cn, smeng@iphy.ac.cn, mliu@iphy.ac.cn*

**ABSTRACT:**

We introduce a rapid, accurate framework for computing atomic migration barriers in crystals by combining universal machine-learning force fields (MLFFs) with 3D potential-energy-surface sampling and interpolation. Our method suppresses periodic self-interactions via supercell expansion, builds a continuous PES from MLFF energies on a spatial grid, and extracts minimum-energy pathways without predefined NEB images. Across twelve benchmark electrode and electrolyte materials—including $LiCoO_2$, $LiFePO_4$, and LGPS—our MLFF-derived barriers lie within tens of meV of DFT and experiment, while achieving $\sim 10^2 \times$ speedups over DFT-NEB. We benchmark GPTFF, CHGNet, and MACE, show that fine-tuning on PBE/PBE+U data further enhances accuracy, and provide an open-source package for high-throughput materials screening and interactive PES visualization.

**Introduction:**

Diffusion is one of the most pervasive and significant phenomena in nature, underlying a vast array of natural and industrial processes [1]. Its influence spans numerous applications, including lithium-ion batteries, fuel cells, catalytic reactions, and alloy formation—all of which rely fundamentally on this mode of mass transport.

For example, in lithium-ion batteries, the diffusion of lithium ions between the anode and cathode is crucial for both charging and discharging, directly affecting battery capacity and lifespan [2]. In fuel cells, diffusion strongly impacts the performance and durability of cell materials; in proton exchange membrane fuel cells (PEMFCs), the transport of gases such as oxygen, carbon monoxide, and hydrogen sulfide through the membrane governs efficiency and operational stability [3]. In the steel industry, the high diffusivity of hydrogen in α-Fe can lead to hydrogen embrittlement, adversely affecting the mechanical properties and longevity of steel [4]. Within cellular biology, the diffusion of amino acids and nutrients is vital for cellular function and metabolism [5]. As a fundamental mechanism of mass transport, diffusion not only shapes scientific research but also holds profound significance in diverse fields, forming the basis for countless natural phenomena and technological advancements.

During atomic migration, ions must overcome an energy barrier—known as the migration barrier or activation energy—which governs the ease of ion movement and, consequently, determines a material's ionic conductivity. Thus, accurately evaluating or predicting diffusion within a system is of critical importance.

However, experimental measurement of atomic migration barrier at the atomistic level remains fundamentally challenging. Most available experimental approaches, such as nuclear magnetic resonance (NMR) spectroscopy [6] for tracing specific species or electrical impedance spectroscopy [7] for probing ionic conductivity, rely on detecting the collective behavior of large populations of atoms. These techniques provide macroscopic averages, which often fail to isolate the intrinsic migration rates of individual atoms. As a result, experimentally determined migration rates can significantly deviate—sometimes by 4 to 5 orders of magnitude [2]—from the actual microscopic migration speeds, due to the inability to separate atomic-scale processes from the complex, composite signals present in real materials.

This limitation highlights the value of computational methods, such as the nudged elastic band (NEB) [8] and drag methods, which can directly probe atomic migration mechanisms on a microscopic scale. Such simulations not only complement experimental data but also enable the prediction of diffusion characteristics before experiments are conducted, thereby offering a powerful tool for the rational design and discovery of new materials.

However, current methods for estimating atomic migration energy barriers still have notable limitations. For instance, the Nudged Elastic Band (NEB) method accurately identifies the minimum energy pathway and transition state structure by optimizing a series of intermediate "images" between the initial and final atomic configurations, thus mapping the most probable diffusion route. While effective, NEB and related approaches—such as the Climbing Image NEB (CI-NEB) [9] developed by Henkelman et al.—are heavily dependent on density functional theory (DFT) calculations, making them computationally demanding. Similarly, Ab Initio Molecular Dynamics (AIMD) [10] simulations can capture collective ion migration phenomena, such as concerted diffusion, but also entail significant computational costs.

On the other hand, empirical molecular dynamics (MD) and bond electrostatic-related approaches, such as the bond valence method [11], are generally too inaccurate to provide reliable migration barriers, further limiting their practical applications. As such, there is an urgent need for improved methods that can accurately and efficiently estimate atomic migration energy barriers in solids.

In this paper, we present a novel approach for the rapid and accurate calculation of atomic diffusion in crystals. By employing universal machine learning force fields (MLFFs), our method enables minute-scale estimations of atomic diffusion in a given crystal, achieving accuracy comparable to DFT but with a computational speed that is approximately $10^2$ times faster. This represents a substantial improvement over traditional empirical force fields. To support this method, we are also releasing an open-source software package, namely FastTrace, capable of identifying migration pathways, visualizing potential energy surfaces, and calculating migration barriers. Our approach and accompanying program greatly improve computational efficiency, making them well suited for high-throughput screening of next-generation electrode and electrolyte materials. Furthermore, the capability to visualize potential energy surfaces provides valuable insight into the microscopic mechanisms governing ion migration.

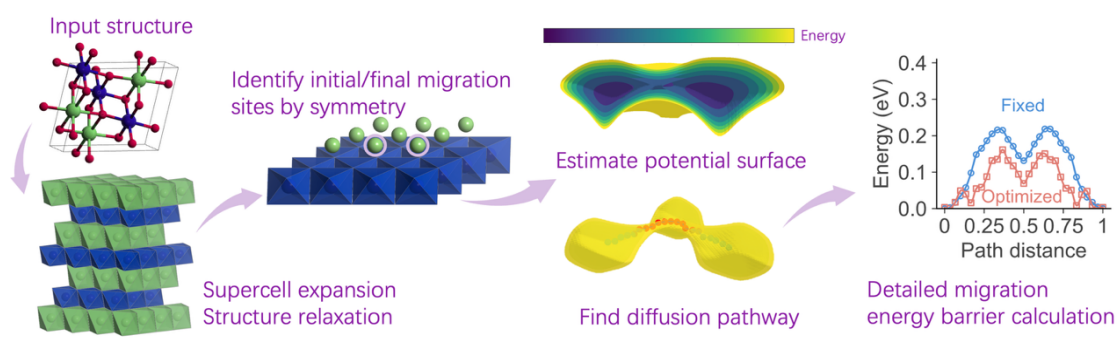

Figure 1. The operational workflow of FastTrack. Taking layered $LiCoO_2$ as an example, Li and Co atoms are shown in green and blue, respectively. The program outputs a three-dimensional visualization of the equipotential surface, the migration pathway (in red), and the corresponding migration energy barrier profile.

The operational workflow of our software package is depicted in Figure 1. First, the input crystal structure undergoes a rational supercell expansion to minimize interactions between the migrating ion and its periodic images, which arise from periodic boundary conditions. Since machine learning force fields generally employ a finite cutoff radius—typically around 6 Å for atomic interactions—we expand the supercell until all lattice vectors exceed 7 Å. This ensures that the migrating ion does not directly interact with itself. Following this expansion, the structure is relaxed using MLFFs to obtain a stable starting configuration. Next, symmetry operations are applied to systematically identify all symmetrically distinct initial and final states relevant to atomic migration. Additionally, the program analyzes the local ionic environment to provide migration pathway suggestions; for instance, it recognizes that single vacancies and divacancies in layered LiCoO$_2$ correspond to different migration mechanisms.

Then, a void space is a created within the structure by removed certain amount atoms to allow the diffusion to happen. This step should be done by exactly mimic the real-world scenario as the diffusion can be viewed as either the hopping of an atom or, in another point of view, the moving of the void space. By creating the void space, as shown in this paper, it is then possible for us to scan the energy potential surface within the space, and then determining the location the initial and final migration sites and searching the diffusion path along the energy potential surface.

Here, we assume that the diffusion mechanism is attributed to the move of a single atom, hence some of the other diffusion pictures such as the concerted diffusion are not considered as in this version of method, but they indeed can be treated similarly in a slightly more complicated version. In our code/method, we do include the three-atom void for the layered structure as it was discovered by existing paper as that single vacancies and divacancies in layered LiCoO$_2$-like structure can show significantly difference diffusivities, and the LiCoO$_2$-like are a common structure for cathode materials in secondary batteries hence people study the diffusivity of them alot.

Then once the void is created, the machine learning force fields (MLFFs) is employed to qucikly scan the energy potential surface of the void by calculation the energy chance by moving one diffusive atom in the void and scan the entire void in the grid fashion. This procedure produces a set of energies as a function of the atom's coordinates (x,y,z). We, then, apply a mathematical interpolation and smoothing operation to the energy potential surface to ensure that the diffusion path can be estimated nicely.

A dedicated algorithm is applied to identify the energy-favoerable migration path through the void. A notable advantage of our approach is the ability to sample migration pathways with high spatial resolution, reducing the likelihood of missing critical features such as saddle points. Once the diffusion path is identified, the miguration energy barrier can be quicly obatined too. This contrasts with the traditional NEB method, in which the diffusion pathway is started from a mostly a guess, and missing a reasonable initial diffusion pathway can lead to unreliable calculation result. Our code provide the the energy potential surface as the output, therefore, once can use the output file to plot and visulazing the energy potential surface.

Considering that that interaction between the moving atom and the structure framework, and during the diffusion process the atoms next to the moving atom can feel the drug and repulsion and in turn move a little bit. We added an extract process in the diffusion calculation workflow to relax the atoms within a certain cutoff radius around the migrating ion to obtain a more accurate migration barrier, which more closely approximates the real situation. We performed tests for the effective interaction radius of atoms in the relaxed structure and uses 2.8 Å as the cutoff radio for relaxation throughout this paper, otherwise stated. The magnitude of the energy difference between the two migration barriers also reflects the structural stability.

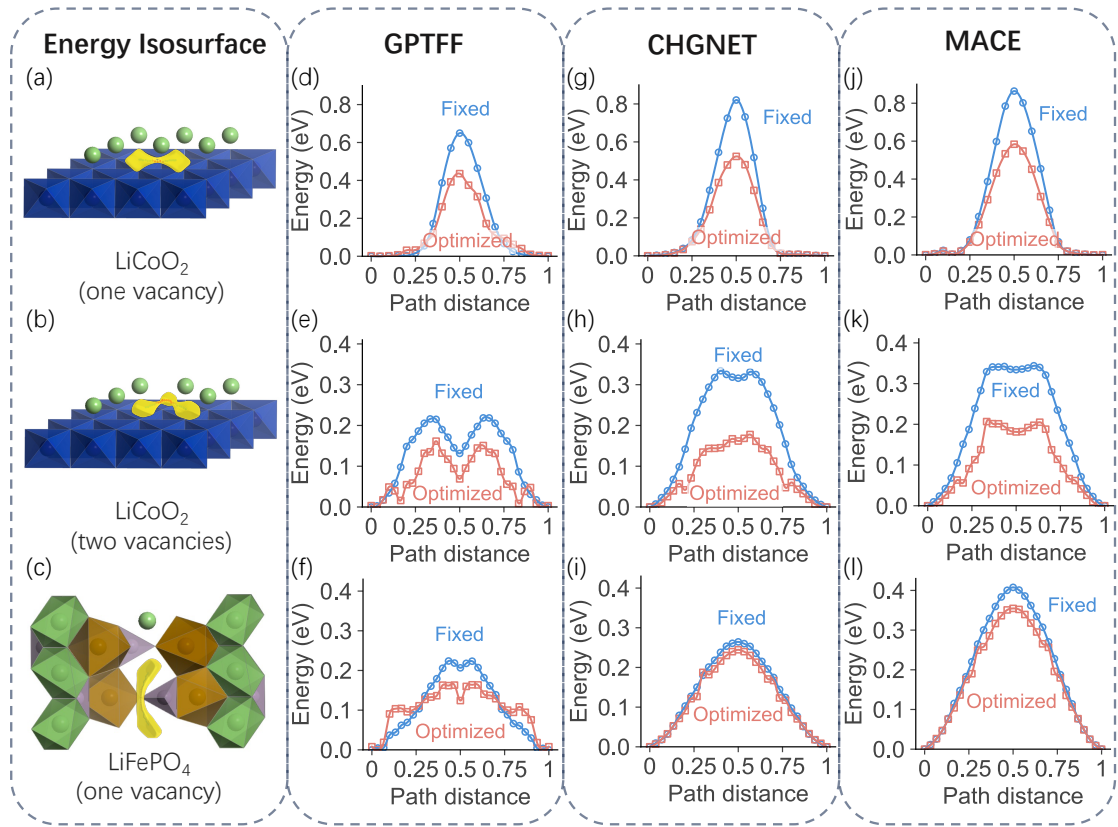

Figure 2. (a-c) Calculated potential energy surfaces (PES) for Li-ion migration. the yellow isosurface marks the region that lies 100 meV above the saddle-point energy.(d-l) Migration energy barrier profiles calculated using different MLFFs. The blue curves correspond to results with other atoms fixed, while the red curves correspond to results with partial atomic relaxation.

Figure 2 illustrates two representative applications of our method, beginning with the cathode material layered $LiCoO_2$. Panels (a) and (b) show calculated potential energy surfaces (PES) for Li-ion migration in single-vacancy and divacancy environments, respectively. In each case, the yellow isosurface marks the region that lies 100 meV above the saddle-point energy, offering an intuitive three-dimensional visualization of the migration landscape. In the single-vacancy configuration, Li ions traverse a pathway flanking the midpoint of an O–O bond, yielding a migration barrier of approximately 600 meV. By contrast, under divacancy conditions, the ion moves directly through the center of a tetrahedral void, where the calculated barrier drops to about 250 meV. These values closely match prior NEB results [12], validating both the accuracy and the enhanced sampling resolution of our approach.

We further demonstrate the versatility of our software with LiFePO$_4$, a well-known one-dimensional ionic conductor. As shown in panel (c), the PES reveals a narrow channel along the [010] direction, with a migration barrier near 300 meV. Notably, the difference between fully relaxed and fixed-lattice calculations is negligible, a consequence of the intrinsic rigidity of the PO$_4^{3-}$ polyanion framework. This rigidity not only underpins the minimal lattice distortion during Li migration but also reflects the excellent thermal stability of phosphate-based electrodes. Thus, beyond high-throughput screening for low-barrier diffusion pathways, our method simultaneously flags candidate structures that combine fast ion transport with robust thermal resilience.

We then benchmarked three state-of-the-art universal MLFFs—GPTFF [13], CHGNet [14], and MACE [15] [16]—to evaluate their out-of-the-box performance for ion diffusion in solids. Each model is built upon distinct training datasets and network architectures, yet all share the ability to reproduce DFT-quality forces and energies sufficient fidelity. As shown in figure 3, all the tested MLFFs can be used to quickly evaluate the migration energy barrier in matter of minutes. While absolute barrier heights exhibit minor variations between MLFFs—and even among DFT calculations themselves, depending on pseudopotential choice [17], Hubbard U corrections, initial path selection, and the number of NEB images—our results consistently fall within the range of expected values. In particular, CHGNet and MACE yield nearly identical barriers, reflecting their use of the similar training dataset.

Importantly, the software we are releasing is MLFF-agnostic: any compatible force field can be plugged into the workflow, allowing users to balance accuracy and speed according to their needs. In addition, dispersion corrections can be directly incorporated into the program [18] [19], which is particularly important for layered materials, although this significantly reduces the computational speed. Together, these benchmarks confirm that our interpolation-based approach, when paired with modern MLFFs, provides a robust, high-throughput means to predict diffusion pathways and activation energies across a wide variety of crystalline solids.

To test transferability across chemistries, we applied each MLFF to calculate migration barriers in a set of twelve prototypical electrode and solid-state electrolyte materials, covering diverse cations and anions (Figure 3). Without any system-specific retraining, all three potentials produced barrier estimates in close agreement with reference DFT values, demonstrating their suitability for rapid, high-throughput screening of candidate materials.

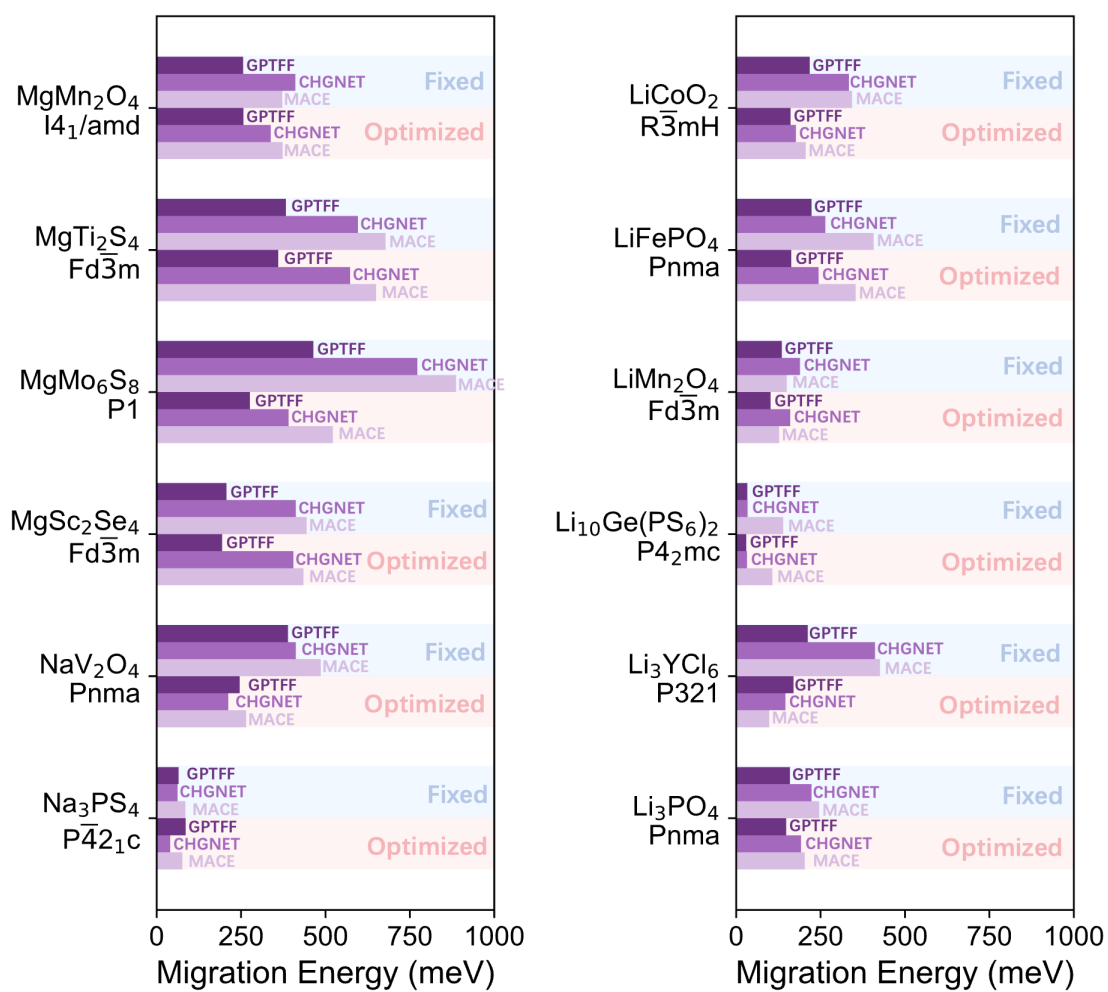

Figure 3. Migration barriers of twelve prototypical electrode and solid-state electrolyte materials. Different shades of purple represent different MLFFs, while the blue and red backgrounds indicate results obtained with fixed atoms and with atomic relaxation, respectively.

To validate our approach, we compared against average Em values reported in the literature for a representative set of materials [17]: $LiCoO_2$ (~230 meV), $LiMnO_2$ and $LiFePO_4$ (~400 meV each), $Li_{10}GeP_2S_{12}$ (LGPS, ~260 meV [20]), $Li_3YCl_6$ (~190 meV [21]), $Li_3PO_4$ (~400 meV), $Na_3PS_4$ (~60 meV), $NaV_2O_4$ (~400 meV), $MgMn_2O_4$ (~700 meV), $MgTi_2S_4$ (~670 meV), $MgSc_2Se_4$ (~370 meV), and $MgMo_6S_8$ (~360 meV) [22]. Our program's predictions lie squarely within these ranges, with only slight underestimation observed for spinel compounds and LGPS. When comparing across force fields, GPTFF tends to yield marginally lower barriers while MACE trends slightly higher, yet all three models produce mutually consistent results. This concordance demonstrates that universal, pre-trained MLFFs offer robust, out-of-the-box accuracy—enabling reliable, high-throughput screening and direct comparison of ionic diffusion properties across diverse crystalline solids.

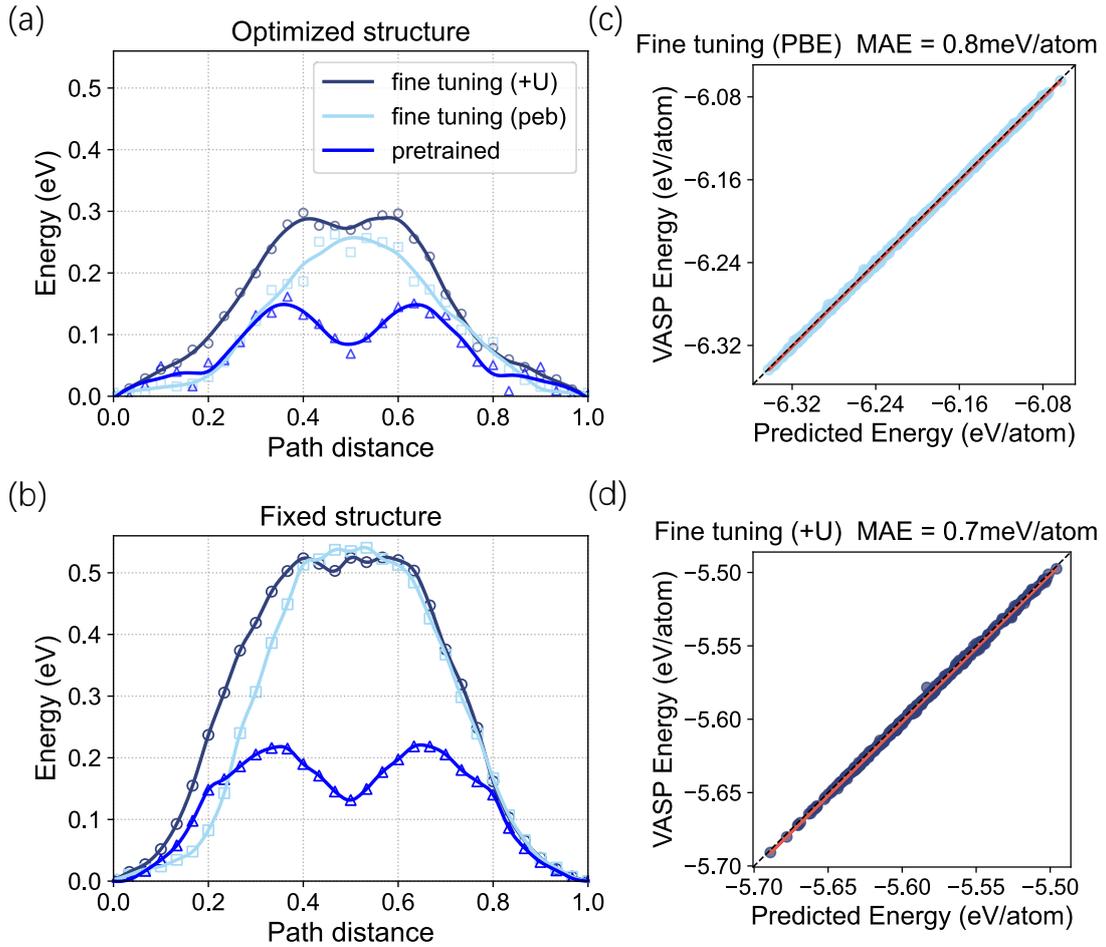

Figure 4. (a) Migration energy barrier profiles for relaxed atoms and (b) for fixed atoms, calculated with different parameters (indicated by lines of different colors). (c) and (d) show the MAE on the test set when trained with PBE and PBE+U data, respectively.

Universal MLFFs, while highly versatile, can exhibit degraded performance on atomic configurations that deviate significantly from their training set, a bias often observed as "potential energy surface softening" [23]. Task-specific fine-tuning offers a direct route to mitigate this effect by enriching the model's representation of transition-state geometries and thereby sharpening the predicted energy landscape. In practice, a fine-tuned MLFF not only improves the fidelity of the potential energy surface but also provides an alternative to conventional NEB calculations for estimating migration barriers.

To illustrate this, we fine-tuned GPTFF on a LiCoO$_2$ dataset generated with both PBE [24] and PBE+U [25] [26] functionals using VASP [27]. As shown in Figure 4, the functional choice markedly influences the computed barriers: both PBE and PBE+U–refined models yield uniformly higher migration energies compared to the original pre-trained force field, reflecting a more accurate sampling of high-energy configurations. These findings underscore the dual importance of training-data coverage and the underlying ab initio methodology in determining MLFF accuracy.

Looking ahead, advances in MLFF architectures, the proliferation of diverse, high-quality training datasets, and continual refinement of first-principles data generation will further enhance barrier predictions within our workflow. As universal force fields become ever more robust, our interpolation-based approach will correspondingly deliver even greater precision and reliability.

**Discussion:**

Machine-learning force fields (MLFFs) are rapidly transforming how we model atomistic processes in solids, offering a compelling complement to traditional density functional theory (DFT). By directly predicting potential energy surfaces, MLFFs can replace many DFT-based tasks—ranging from bond energy calculations to transition-state searches—at a fraction of the computational cost. This paradigm shift parallels developments in computational drug discovery, where rapid, in silico screening of candidate molecules precedes laboratory testing. In our workflow, a single migration-barrier estimate takes only about 20 minutes on a single NVIDIA Tesla A100 GPU(when using GPTFF as the MLFF), versus hours or days for comparable DFT–NEB calculations.

Traditional methods each carry inherent trade-offs. The nudged elastic band (NEB) technique reliably locates minimum-energy paths and saddle points but is hindered by high cost and sensitivity to the chosen initial and final states. Ab initio molecular dynamics (AIMD) captures temperature effects and concerted ion motions yet remains limited by short time and length scales. Kinetic Monte Carlo (KMC) extends those scales but depends critically on accurate input rates and struggles with complex, multi-body interactions. Empirical bond-valence methods (BVSM/BVSE) offer speed but sacrifice quantitative precision. Our interpolation-based MLFF approach addresses many of these challenges: it eliminates the need for predefined endpoints, maintains near-DFT accuracy, and delivers results in orders-of-magnitude less time. Its current limitation lies in fully capturing concerted migrations—a task better suited to AIMD or advanced KMC when collective effects dominate.

One ongoing area of investigation is the treatment of electronic and magnetic effects. DFT–NEB for systems with localized d- or f-electrons can suffer from energy fluctuations between images, especially when electron transfer or spin states change along the pathway. Standard MLFFs typically omit explicit electronic degrees of freedom, focusing instead on atomic forces and energies. An exception is CHGNet, which incorporates magnetic-moment predictions and thus carries implicit electronic information; however, its migration-barrier predictions remain largely consistent with other MLFFs. Whether this electronic sensitivity substantially impacts barrier accuracies for magnetic or charge-transfer processes merits further study.

The choice of functional or dataset used for training MLFFs also plays a critical role. In our benchmarking, barriers predicted by CHGNet and MACE—both trained on PBE+U data—are nearly identical, whereas GPTFF, built on a larger and more diverse dataset, tends to yield slightly lower values. This mirrors the known functional dependence of NEB barriers (e.g., $LiCoO_2$ barriers range from ~200 meV with GGA to ~519 meV with GGA+U [28]). Fine-tuning MLFFs on task-specific, transition-state–rich data can correct for "potential-energy-surface softening" and bring their predictions into even closer alignment with high-level DFT.

Beyond accuracy and speed, our software furnishes interactive visualizations of potential energy landscapes and an efficient algorithm for extracting minimum-energy paths. Users can readily inspect three-dimensional isosurfaces of saddle-point regions and explore alternative diffusion routes, strengthening microscopic insights into ion-migration mechanisms.

We also attempted to compute migration barriers by combining machine learning force fields (MLFF) with the NEB method, which yields a much faster computation too. During manuscript preparation, we became aware of a concurrent study pursuing a similar MLFF–NEB integration [29], and earlier work has applied machine-learning molecular dynamics (MLMD) for high-throughput ionic diffusivity screening [30]. In contrast, our workflow directly generates real-space potential-energy surfaces and delivers exceptional computational efficiency, making it a rapid and robust alternative to both NEB and MLMD approaches.

Looking forward, as MLFF architectures evolve, training datasets expand, and first-principles data generation advances, we anticipate that MLFF-based diffusion modeling will become ever more precise, robust, and ubiquitous in the high-throughput design of next-generation electrode and electrolyte materials.

**Conclusion**

In summary, we have demonstrated a fast, accurate, and flexible framework for predicting atomic diffusion barriers in crystalline solids by integrating universal machine-learning force fields with three-dimensional potential-energy-surface sampling and interpolation. Our approach, which obviates the need for predefined NEB images, delivers migration energy estimates within tens of millielectronvolts of DFT and experimental benchmarks while achieving speedups of roughly two orders of magnitude. Through systematic benchmarking of GPTFF, CHGNet, and MACE—and by fine-tuning on PBE and PBE+U datasets—we have shown that both data coverage and functional choice critically influence barrier accuracy. The accompanying open-source software facilitates high-throughput screening of electrode and electrolyte materials, complete with interactive visualization of energy landscapes and automated pathfinding. Looking ahead, extending our workflow to incorporate dispersion corrections, capture concerted migration events, and leverage next-generation MLFF architectures will further enhance its predictive power and broaden its applicability in the design of advanced energy materials.

## ACKNOWLEDGMENTS

This research is supported by National Key R&D Program of China (Grant No. 2024YFF0508500), Ministry of Science and Technology (Grant No. 2021YFA1400201 and No.2021YFA1400503), National Natural Science Foundation of China (Grant No. 12025407 and No. 12450401), and Chinese Academy of Sciences (Grant No. YSBR047). The computational resources are provided by the Data-Driven Computational Materials Discovery Platform at Songshan Lake laboratory

## DATA AVAILABILITY

Our code will be made publicly available after the publication of the paper.